\documentclass[a4paper]{article}
\usepackage{graphicx}
\usepackage{setspace}

\begin{document}

\title{\LARGE \bf The Nature of Alpha \footnote{
An expanded version of this paper was published (and should be used for citations) as: \newline Arthur M. Berd, ``Investment Strategy Returns: Volatility, Asymmetry, Fat Tails and the Nature of Alpha'', in {\em Lessons from the Financial Crisis}, A.M.\ Berd (ed.), RiskBooks, 2010}}
\author{ {\Large \bf Arthur M. Berd} \\
BERD LLC}
\date{February 2011}
\maketitle

\begin{abstract}
We suggest an empirical model of investment strategy returns which elucidates the importance of non-Gaussian features, such as time-varying volatility, asymmetry and fat tails, in explaining the level of expected returns. Estimating the model on the (former) Lehman Brothers Hedge Fund Index data, we demonstrate that the volatility compensation is a significant component of the expected returns for most strategy styles, suggesting that many of these strategies should be thought of as being `short vol'. We present some fundamental and technical reasons why this should indeed be the case, and suggest explanation for exception cases exhibiting `long vol' characteristics. We conclude by drawing some lessons for hedge fund portfolio construction.
\end{abstract}

\newpage

\section{Introduction} \label{berd:sec:intro}

The key question in investment management is to understand the sources of investment returns. Without such understanding, it is virtually impossible to succeed in managing money. In this article, we explore, using an empirical model of hedge fund strategy returns, the importance of non-Gaussian features, such as time-varying volatility, asymmetry and fat tails, in explaining the level of expected returns. We demonstrate that the volatility compensation is often a significant component of the expected returns of the investment strategies, suggesting that many of these strategies should be thought of as being `short vol'. The notable exceptions are the CTA strategies and certain fixed income and FX strategies. We suggest a fundamental explanation for this phenomenon and argue that it leads to important adjustments in capital allocation.

The experience of the past few years has largely confirmed our hypothesis, laid out originally in conference presentations \cite{Berd-NatureOfAlpha-2008,Berd-VolStrategies-2009} and earlier strategy papers \cite{Berd-Tetyevsky-2002,Berd-VolCycle-2007}. In particular, we believe that the Quant Crunch experience in August 2007, and the widespread large losses in September 2008 till March 2009 period, followed by large gains after March 2009 till early 2010 period, are broadly in line with the volatility dependence of strategy returns presented in this paper and as such are neither exceptional not surprising. 

Moreover, the increased correlation across most strategies during this period can be largely attributed to the common driving factor which is the level of volatility. Knowing that most strategies' returns are highly directionally dependent on the volatility, and having a lot of prior empirical evidence that the volatility spikes across markets are highly correlated, leads one to a straightforward conclusion that the correlation across strategies will also rise during a market-wide spike in risk.

There are many theories and plausible hypotheses about the driving factors explaining the returns. The Capital Asset Pricing Model \cite{Sharpe-CAPM-1964,Lintner-CAPM-1964,Mossin-CAPM-1964}, which postulates a specific relationship between the expected excess returns of assets and their co-variation with market returns, is among the pillars of modern financial theory. A more elaborate and pragmatically more satisfying Arbitrage Pricing Theory \cite{Ross-APT-1976} expands on CAPM insights by adding more explanatory variables and allowing more flexibility in defining the common driving factors. The popular Fama-French framework \cite{Fama-French-1993} can be seen as a particularly successful example of APT in application to stock returns.

Majority of these theories are focused on explaining the returns of tradable assets, such as stocks, bonds or futures. While this is clearly the most granular level which is of interest to the researchers, we believe there is much to be gained by focusing instead on returns of typical investment strategies. Indeed, over the past decades, many such strategies, from quantitative long/short equity investing to CTA and to convertible arbitrage, have become well-established and boasting must-have allocations in most alternative investment portfolios.

We can take this argument even further, stating that not only it is possible to model the strategy returns without knowing its composition, but it is actually important to do that. Suppose we had the benefit of knowledge of some particular hedge fund's portfolio composition at a given point in time. Would that give us a better understanding of the nature of its returns over time? The answer, of course, is -- it depends. If the strategy is slow and with low turnover, then yes -- the current composition does provide important clues for the future returns. However, if the strategy has either high turnover or potentially large trading activity driven by market events, then the current composition can actually be quite misleading for predicting the future returns.

Let's consider two simple cases. In the first, the strategy at hand is what used to be called a {\em portfolio insurance}, which in essence was trying to replicate the downside S\&P 500 put option position by dynamically trading in the S\&P 500 futures. Clearly, this strategy has a pretty high turnover, driven by the daily fluctuations of the S\&P 500 index. Moreover, its return pattern, by design, matches not that of a linear futures position, but that of a non-linear option position. In particular, the strategy is supposed to make money if the market fluctuates strongly but returns roughly to the same level over time, while the linear futures position of course would have close to zero return in this case.

In the second example, consider a simplified macro strategy, where the portfolio manager switches between the long and short positions in some macro index, such as the very same S\&P 500 futures, depending on a bullish/bearish signal for the economy, e.g.\ based on some macroeconomic model. These signals, by construction, would not change very frequently, sometimes staying in one state for many months or even years. But when they do, the impact on the strategy will be dramatic -- it will reverse the sign of the returns. So, if one is interested in the distribution of the returns of such strategy over long periods of time encompassing the business cycle, they will be quite different from the distribution of the underlying S\&P 500 returns, even if for a typical month the returns might actually be identical.

So, it appears that in order to estimate the returns characteristics of the strategy, it is less important to know {\em what} does the strategy trade, and more important to know {\em how} it trades. Moreover, majority of professional investors who actually produce these returns exhibit a good level of discipline in what they do. Depending on their style and competitive advantage, they typically stick to certain patterns of behavior and are driven by relatively stable methodologies which they use for estimating risks and returns within their universe of tradable securities, and for allocating capital across particular investments. It is these behavior patterns and stable methodologies that leave their imprint on the corresponding strategy returns. And while this statement is obviously true with respect to so called {\em quantitative} or {\em systematic} investment managers, we think it largely applies even to investors following the {\em discretionary} style.

One could take a step further and say that each of the investment funds can be considered an independent company, whose stock (NAV) performance reflects its business model and the management style. Unlike bricks and mortar businesses, such `virtual' companies do not have assets in which their business model is ingrained. And looking at their balance sheet produces little more than the knowledge of their leverage and other basic facts. Unlike Gilette or Starbucks, you can't really judge these companies by what products or services they produce, or what factories and stores they have. And while in real companies the management style also sometimes matters a lot (think Apple and Steve Jobs, or GE and Jack Welch), in the virtual financial holding companies that the investment funds really are, the management style and methodology is really the only thing that matters.

To discern meaningful patterns in such nebulous things as management style and methodology one must start by modeling the strategy returns. There are three basic ways to model them, in order of decreasing granularity:

\begin{description}
\item[Micro-replication:] In this approach, one actually tries to build a systematic process similar to each particular strategy in complete detail including specific trading signals in variety of chosen instruments, and then fine tunes a few parameters to match the observed returns of the investment strategy benchmark, such as a particular hedge fund index. 
\item[Macro-replication:] In this approach, one tries to figure out which macro variables influence the strategy's returns, and tries to build a time-series forecasting model with these variables. 
\item[Parametric:] In this approach, the strategy return time series is modeled endogenously, in a manner similar to modeling `elemetary' asset returns, e.g.\ by well-known econometric methods. 
\end{description}

The parametric approach, following the pioneering work of Fung and Hsieh \cite{Fung-Hsieh-1997}, attempts to simply understand the properties of hedge fund returns, while the replication approaches (see \cite{Takahashi-Yamamoto-2010} for a recent review) attempt to model the hedge fund strategies themselves in their full dynamics, either in a bottom-up or top-down manner. Each of the approaches has its pros and cons. 

Micro-replication \cite{Agarwal-Naik-2000,Wallerstein-Tuchschmid-Zaker-2010} has been successful in mimicking some mainstream hedge fund strategies, such as CTA, equity stat arb, or merger arb, so much so that there exist ETFs and ETNs making such strategies available for broad investor audience. But for many other types of strategies, such an approach is hopelessly difficult and ambiguous.

The macro-replication is somewhat more universally applicable, especially in strategies which are known as {\em alternative beta}, i.e.\ where the performance of the strategy is driven by its exposure to well-identified market risk factors \cite{Fung-Hsieh-2002,Fung-Hsieh-2004,Hasanhodzic-Lo-2005}. Here too, certain alternative beta strategies have been sufficiently popular to launch a widely marketed ETF or ETN, for example iPath Optimized Currency Carry ETN (ICI) and iPath S\&P 500 VIX Short-Term Futures ETN (VXX). Such an approach, by design, does not explicitly model the strategy's {\em pure alpha}, treating it as a residual constant return. It also assumes that the strategy maintains a constant exposure to the chosen set of macro factors, which is also not necessarily a universally valid assumption. 

On the other hand, the parametric approach, which we advocate in this paper, does not concern itself with the detailed composition of returns or indeed with their attribution to observable market factors. Instead, it treats the investment strategies in a holistic manner, as opaque financial assets with little more than their net asset values (NAV) and returns visible to outside observers. It is universally applicable to any strategy that one may consider modeling, without complicating the analysis with additional assumptions. 

The main question here is the stability and interpretability of the results. The way to attain a positive answer to this question is not to modify the econometric model or make it more complex, but to choose a well-designed set of benchmark indexes or fund peer groups for estimation. This is akin to an old approach of modeling all stocks on the basis of their price/earnings (PE) ratios, but recognizing that companies from different industry sectors or with widely different market caps may have substantial differences in the manner in which their PE impacts their future returns. There, too, the simple solution is to divide the universe of all stocks into relatively uniform peer groups, and to only compare the PE within the same group.

Thus, to get sensible results from our chosen parametric approach, we must apply it to a set of investment strategy indexes which we believe have been constructed in a sufficiently uniform manner and with the appropriate amount of granularity. This requirement has led us to select the Lehman Brothers Hedge Fund Index, which was a part of the overall suite of global indexes built and maintained by Lehman's index group, and benefited from the thorough and disciplined rules-based methodology in the same manner as their better known US Aggregate Bond Index. In our opinion, this set of indexes was much more complete and much better designed than the more widely disseminated CS Tremont, Hennessee or HFRI Indexes. While the latter ones have been around for longer and have possibly more hedge funds in their coverage, they use classification schemes which are outmoded and do not correspond to actual segmentation of the hedge fund universe by investment style or product focus. The Lehman Brothers Hedge Fund Index had, in contrast, a full set of available sub-indexes classified by style, product or region, all constructed in their typical consistent fashion. Unfortunately for the analysts, this index product did not survive the demise of the parent company and was discontinued by Barclays Capital in early 2009. Still, it offered a consistent set of data from early 2000 until the end of 2008, a period that saw two recessions, two market crashes and a multi-year boom, and so it appeared to be still the best choice for our research purposes, despite not being available post 2009.

\section{An Econometric Model of Strategy Returns} \label{berd:sec:model}

In this section, we specify the econometric model of investment strategy returns. It is a straightforward generalization of the celebrated GARCH family of models \cite{Engle-ARCH-1982,Bollerslev-GARCH-1986,Engle-Bollerslev-Nelson} designed to capture the well-known stylized facts regarding the dynamics of financial time series, such as the clustering and mean reversion of time-varying volatility, fat-tailed distribution of periodic returns, and asymmetric volatility responses which serve as a dynamic mechanism of generation of non-Gaussian features of long-run aggregate returns. Given that some investment strategies exhibit not only long term asymmetries but also quite visible short-term asymmetries, we expand the definition to also include asymmetric fat tails of periodic returns. Finally, as it is the primary objective of our study, we allow non-zero conditionally time-varying means of the periodic returns. 

We call this setup a Generalized Asymmetric Autoregressive Conditional Heteroscedasticity (GAARCH) model \cite{Berd-NatureOfAlpha-2008,Berd-VolStrategies-2009}. The GAARCH(1,1) model includes a single lag for both past returns and past conditional volatilities:
\begin{eqnarray} 
r_{t} & = & \mu_{t} + \sigma_{t} \epsilon_{t}  \label{gaarch_process} \\
\mu_{t+1} & = & \alpha + \gamma \sigma_{t+1}^2 \label{gaarch_cmean} \\
\sigma_{t+1} & = & \sigma_0 \cdot \sqrt{1 + \chi_{t+1}} \label{gaarch_cvol} \\
\chi_{t+1} & = & \left[m_{2}^{-} \eta^{-} + m_{2}^{+} \eta^{+} + \beta \right] \chi_{t} \nonumber \\
& + &  \eta^{-} \left(1 + \chi_{t}\right) \left(\epsilon_{t}^2 \cdot 1_{\epsilon_{t} < 0} - m_{2}^{-} \right) \nonumber \\
& + &  \eta^{+} \left(1 + \chi_{t}\right) \left(\epsilon_{t}^2 \cdot 1_{\epsilon_{t} \geq 0} - m_{2}^{+} \right)  \label{gaarch_cxvar} \\
m^{-}_{2} & = & E\left\{\epsilon^2 \cdot 1_{\epsilon < 0}\right\}  \label{gaarch_trunchmoments-} \\
m^{+}_{2} & = & E\left\{\epsilon^2 \cdot 1_{\epsilon \geq 0}\right\} \label{gaarch_trunchmoments+}
\end{eqnarray}

\noindent Here, we used the following notations:

\begin{itemize}
	\item $r_{t}$ is the single-period (in our case, monthly) return of the asset.
	\item $\epsilon_{t}$ are the residual returns for the period ending at $t$, which are i.i.d.\ variables with zero mean and unit variance.
	\item $\mu_{t}$ is the conditional mean of asset returns for the period ending at $t$.
	\item $\sigma_0$ is the unconditional volatility of asset returns.
	\item $\sigma_{t}$ is the conditional volatility of asset returns for the period ending at $t$.
	\item $\chi_{t}$ is the conditional excess variance, equal to the percentage difference between the conditional and unconditional variance $\chi_{t} = \sigma_{t}^2/\sigma_0^2 - 1$.
	\item $m^{\pm}_{2}$ are the truncated upside and downside second moments of single-period residual returns.
\end{itemize}

The specification of the GAARCH model incorporates a scale-invariant reparameterization of the standard stationary GARCH model \cite{Engle-ARCH-1982,Bollerslev-GARCH-1986}, separating the level of the unconditional volatility $\sigma_0$ and the conditional excess variance $\chi_{t}$ process. The dynamic asymmetry of volatility is specified in a manner similar to TARCH \cite{Zakoian-1994} (see also GJR-GARCH \cite{GJR-1993}), but we redefined the ARCH terms in a more symmetric fashion, without specifying which of the signs (positive or negative) is more influential. This is because in some asset classes, notably credit and volatility, the upside shocks are more influential, while in others, like equities or commodities, the downside shocks are more influential. The symmetric specification is obviously equivalent to TARCH but allows one to have a more natural positivity restriction on ARCH coefficients, if so desired. 

The conditional mean specification in (\ref{gaarch_cmean}) is more or less in line with the conventional APT assumptions, if we assume that the strategy is uncorrelated with the overall market and that variability of returns is priced as an alternative beta. The unconditional mean $\alpha$ can be considered the `true alpha' of the strategy, i.e. its excess return above the compensation for systematic risks that the strategy takes. 

The conditional mean process can also be rewritten in a form that allows a more subtle attribution of expected returns. Introducing the parameter $\Gamma = \gamma \sigma_0^2$, which we will call `convexity compensation' because it resembles the additional return that any convex investment acquires under fair pricing rules, and using the definition of the excess variance process $\chi_{t}$, we get:

\begin{equation} \label{gaarch_cmean_parts}
\mu_{t} = \alpha + \gamma \sigma_0^2 + \gamma \sigma_0^2 \left( \frac{\sigma_{t}^2}{\sigma_0^2} - 1 \right) = \alpha + \Gamma + \Gamma \chi_{t} 
\end{equation}

This equation can be interpreted as a three-way attribution of conditional expected returns to the true alpha $\alpha$, the convexity compensation $\Gamma$, and time-varying excess variance compensation (the latter has an unconditional expected value of zero). One can also add the first two constants to get the {\em risk adjusted alpha} of the strategy:

\begin{equation} \label{gaarch_net_alpha}
\hat{\alpha} = \alpha + \Gamma 
\end{equation}

Finally, let us specify the distribution of the return residuals $\epsilon_{t}$. Our main criteria are that it must be parsimonious (i.e.\ have no more than two free parameters to describe the asymmetric fat tails of a standardized distribution with zero mean and unit variance), and that the estimation of these parameters should be robust with respect to the deviation of sample returns from the zero mean assumption. The latter criterion is necessary because we would like as much separation as possible between the estimation of the GAARCH conditional mean and the estimation of the distribution of residuals. 

The requirement of robustness with respect to sample mean suggests that we should consider distributions whose asymmetry is defined by their tail dependence, rather than by introducing a third order skewness term which could be influenced by the estimate of the mean. From a variety of well-known skewed fat-tailed distributions, Jones and Faddy \cite{Jones-Faddy-2003} skewed t-distribution fits our criteria best. The probability density and cumulative density functions of this distribution are:

\begin{equation} \label{jf_skewt_pdf}
f\left(x \left| \nu^{-}, \nu^{+} \right. \right) 
 = \frac{\left(1 + \frac{x}{\sqrt{\nu + x^2}} \right)^{\frac{\nu^{-} + 1}{2}} \; 
 \left(1 - \frac{x}{\sqrt{\nu + x^2}} \right)^{\frac{\nu^{+} + 1}{2}}}{ 2^{\nu-1} \, \sqrt{\nu} \, B\left(\frac{\nu^{-}}{2}, \frac{\nu^{+}}{2} \right)} 
\end{equation}

\begin{equation} \label{jf_skewt_cdf}
F\left(t \left| \nu^{-}, \nu^{+} \right. \right) 
 = I\left(\frac{1}{2} \left(1 + \frac{t}{\sqrt{\nu + t^2}}\right), \frac{\nu^{-}}{2}, \frac{\nu^{+}}{2} \right)
\end{equation}
where $\nu = \left(\nu^{-} + \nu^{+} \right)/2$, $B(a,b)$ is the beta function, and $I(z,a,b)$ is the incomplete beta function. 

The parameters $\nu^{-}$ and $\nu^{+}$ have the meaning of the left and right tail parameters, as can be seen from the asymptotics of this distribution.

\begin{eqnarray} \label{jf_skewt_asymptotics}
f\left(x \left| \nu^{-}, \nu^{+} \right. \right) & \rightarrow & \left|x\right|^{- \nu^{-} - 1} \; \mbox{for} \; x \rightarrow -\infty \nonumber \\
f\left(x \left| \nu^{-}, \nu^{+} \right. \right) & \rightarrow & \left|x\right|^{- \nu^{+} - 1} \; \mbox{for} \; x \rightarrow +\infty 
\end{eqnarray}

The distribution (\ref{jf_skewt_pdf}) becomes equivalent to the conventional Student's t-distribution when $\nu^{-} = \nu^{+}$, which in turn nests the Gaussian one when $\nu \rightarrow +\infty$, and therefore our entire model specification allows for the goodness of fit comparisons between different models and for the likelihood ratio tests of statistical significance of the estimated parameters.

\section{The Taxonomy of Patterns in Hedge Fund Strategy Returns} \label{berd:sec:taxonomy}

In this Section, we will try to classify the dependence of the quantitative characteristics of hedge fund strategy returns, seen through the empirical fit of the GAARCH model, on the type of peer group and other qualitative characteristics. We specifically highlight the non-Gaussian features of the returns, including the presence of fat tails, the asymmetry between the upside and downside tails, the volatility level and its dynamics, and in particular the asymmetry of volatility response to return shocks (asymmetric leverage effect).

From \cite{Drost-Nijman,Berd-Engle-Voronov-2007} we know that the relative order of importance of these characteristics from the perspective of description of intermediate/long term distribution of returns in financial time series is as follows:

\begin{itemize}
	\item Persistence and mean reversion of vol (GARCH terms) - governs the behavior of conditional volatility and the vol of vol at intermediate timescales.
	\item Asymmetric volatility response (TARCH terms) - is the leading contributor to the non-Gaussian features (tails and asymmetry) of long-run aggregated returns.
	\item Fat-tailedness and asymmetry of periodic returns (non-Gaussian shocks) - is important for short term (up to several periods)
\end{itemize}

We will confine the discussion in this Section to the dynamic properties of the returns, and will postpone the discussion of the conditional means (expected returns) till the next Section.

We apply the GAARCH model specified in Section \ref{berd:sec:model} to the Lehman Brothers Hedge Fund Index (LBHF) and its subindexes. The universe of funds in the index has grown from 325 in the beginning of 2000 to the peak number of 2288 by August 2008, before dropping back to 1583 by early 2009. Given the limited historical monthly data for these indexes, available only from January 2000 until January 2009, and given the extreme realizations of the returns and volatilities in the second part of 2008, which would have dominated the small dataset and skewed the estimation of parameters significantly, we have chosen a somewhat shorter period of January 2000 until April 2008 as our sample period for estimation. 

To ensure a better diversification of the subindex returns, we used the equal-weighted version of the LBHF Index. While this may limit the applicability of our results from the perspective of investability of the corresponding representative indexes, using the AUM-weighted version would have subjected our data sample to dominance of a small number of very large funds, making the corresponding index returns much more idiosyncratic. The equal-weighted version returns, on the other hand, appear quite systematic, and therefore we are able to obtain reasonable model fits despite the limited length of time series. 


The results of the model fit for different size buckets of the LBHF Index are shown in Table \ref{berd:tab:size}. We also report the results of the model fit for style and asset class subindexes of the LBHF Index in Tables \ref{berd:tab:style} and  \ref{berd:tab:asset}, respectively. In each table, we indicate the hierarchy of the subindexes by tabbing, e.g.\ the CTA Trend-following subindex within the CTA index is indicated by an extra tab level. 

For the ease of comparison with well-known industry metrics, we report the values of risk adjusted alpha $\hat{\alpha}$, true alpha $\alpha$, convexity compensation $\Gamma$ and unconditional volatility $\sigma_0$ in annualized percentage points terms. The coefficients which drive the volatility dynamics of the model are reported in GAARCH convention, including the downside $\eta^{-}$ and upside $\eta^{+}$ ARCH coefficients, and the GARCH coefficient $\beta$. And finally, we report the left $\nu^{-}$ and right $\nu^{+}$ tail degrees of freedom of the Jones-Faddy skewed Student's t-distribution of the residual returns.

\begin{table}[tp]
\hspace*{-1cm} \begin{tabular}{|l||c||c|c||c|c|c|c||c|c|} \hline
Index & \multicolumn{3}{c||}{Expected Returns} & \multicolumn{4}{c||}{Volatility Dynamics} & \multicolumn{2}{c|}{Tails} \\ \hline 
 			& $\hat{\alpha}$ & $\alpha$ & $\Gamma$ & $\sigma_0$ & $\eta^{-}$ & $\eta^{+}$ & $\beta$ & $\nu^{-}$ & $\nu^{+}$ \\ \hline \hline
All Funds 												& 8.25 & 16.05 & -7.80 & 5.30 & .26 & .03 & .70 & 15.24 & 16.95 \\ \hline
Fund Size Greater than 50MM   		& 7.86 & 15.20 & -7.34 & 5.23 & .31 & .07 & .65 & 15.50 & 15.82 \\ \hline
Fund Size Greater than 100MM   		& 7.32 & 15.23 & -7.91 & 5.05 & .29 & .11 & .69 & 19.40 & 19.70 \\ \hline
Fund Size Greater than 250MM   		& 7.19 & 14.21 & -7.01 & 4.70 & .30 & .15 & .69 & 25.41 & 29.75 \\ \hline
Fund Size Greater than 500MM   		& 6.98 & 13.59 & -6.61 & 4.83 & .29 & .19 & .56 & 34.89 & 200. \\ \hline
Fund Size Greater than 1 billion 	& 6.68 & 12.28 & -5.60 & 4.31 & .39 & .47 & .44 & 48.78 & 200. \\ \hline

\end{tabular}
	\caption{Estimation of GAARCH model for LBHF size subindexes}
	\label{berd:tab:size}
\end{table}

\begin{table}[tp]
\hspace*{-2cm} \begin{tabular}{|l||c||c|c||c|c|c|c||c|c|} \hline
Index & \multicolumn{3}{c||}{Expected Returns} & \multicolumn{4}{c||}{Volatility Dynamics} & \multicolumn{2}{c|}{Tails} \\ \hline 
 			& $\hat{\alpha}$ & $\alpha$ & $\Gamma$ & $\sigma_0$ & $\eta^{-}$ & $\eta^{+}$ & $\beta$ & $\nu^{-}$ & $\nu^{+}$ \\ \hline \hline
Macro/Directional 																	& 8.17 & 19.54 & -11.37 &  6.85 & .19 & .02 & .73 & 11.74 & 11.52 \\ \hline
\hspace{0.5cm} Macro 																& 7.77 & 12.95 &  -5.18 &  7.34 & .18 & .14 & .74 &  6.89 & 6.07 \\ \hline
\hspace{0.5cm} Long-Short 													& 9.26 & 22.27 & -13.01 &  7.70 & .21 & .04 & .76 &  7.14 & 9.90 \\ \hline
\hspace{1.0cm} LS Long Only 											 & 16.51 & 23.67 &  -7.16 & 12.69 & .18 & .15 & .77 & 19.67 & 200. \\ \hline
\hspace{1.0cm} LS Long Bias 												& 9.00 & 20.81 & -11.81 &  9.56 & .26 & .   & .75 &  7.81 & 9.05 \\ \hline
\hspace{1.0cm} LS Variable Bias 										& 8.50 & 20.44 & -11.94 &  5.84 & .20 & .10 & .74 & 12.72 & 18.02 \\ \hline
\hspace{1.0cm} LS Short Bias 												&  .37 & -6.03 &   6.40 &  8.57 & .12 & .11 & .81 & 11.77 & 10.50 \\ \hline
CTA 																								& 7.44 & -2.95 &  10.39 & 10.34 & .12 & .09 & .65 & 200.  & 52.06 \\ \hline
\hspace{0.5cm} CTA Fundamental 											& 7.58 &  5.49 &   2.09 &  7.14 & .16 & .06 & .80 & 166.36 & 100.45 \\ \hline
\hspace{0.5cm} CTA Trend-following 									& 8.40 & -6.60 &  15.01 & 13.52 & .11 & .09 & .63 & 200.  & 47.46 \\ \hline
Relative Value 																			& 7.87 & 12.68 &  -4.80 &  3.22 & .32 & .   & .75 & 53.27 & 200. \\ \hline
\hspace{0.5cm} RV Statistical Arbitrage 						& 8.65 & 15.03 &  -6.38 &  4.17 & .21 & .   & .64 & 11.41 & 14.74 \\ \hline
\hspace{0.5cm} RV Special Situations/Event Driven 	& 9.09 & 17.69 &  -8.60 &  4.46 & .16 & .   & .88 & 36.77 & 200. \\ \hline
\hspace{0.5cm} RV Distressed 												& 9.32 & 16.18 &  -6.85 &  5.20 & .15 & .   & .85 & 91.89 & 200. \\ \hline
\hspace{0.5cm} RV Reg D 														& 9.48 & 13.20 &  -3.72 &  4.39 & .18 & .   & .80 & 15.90 & 8.63 \\ \hline
\hspace{0.5cm} RV Merger Arbitrage 									& 7.71 & 12.97 &  -5.26 &  3.99 & .20 & .22 & .71 & 21.45 & 125.73 \\ \hline
\hspace{0.5cm} RV Market Neutral-Equity Only 				& 5.33 &  7.37 &  -2.04 &  3.23 & .18 & .20 & .73 &  4.85 & 5.99 \\ \hline
\hspace{0.5cm} RV Broad Relative Value 							& 8.30 & 13.09 &  -4.79 &  2.67 & .37 & .   & .74 & 39.96 & 153.00 \\ \hline
\end{tabular}
	\caption{Estimation of GAARCH model for LBHF style subindexes}
	\label{berd:tab:style}
\end{table}

\begin{table}[tp]
\hspace*{-2cm} \begin{tabular}{|l||c||c|c||c|c|c|c||c|c|} \hline
Index & \multicolumn{3}{c||}{Expected Returns} & \multicolumn{4}{c||}{Volatility Dynamics} & \multicolumn{2}{c|}{Tails} \\ \hline 
 			& $\hat{\alpha}$ & $\alpha$ & $\Gamma$ & $\sigma_0$ & $\eta^{-}$ & $\eta^{+}$ & $\beta$ & $\nu^{-}$ & $\nu^{+}$ \\ \hline \hline
Multi-Market 																		& 8.29 	& 16.94 &  -8.65 & 5.28 & .18 & . & .72 & 129.44 & 200. \\ \hline
\hspace{0.5cm} MM Macro/Directional 						& 8.69 	& 16.14 &  -7.45 & 7.15 & .19 & .02 & .73 & 100. & 83.86 \\ \hline
\hspace{0.5cm} MM Relative value 								& 8.11 	& 17.64 &  -9.54 & 4.01 & .19 & . & .86 & 67.51 & 200. \\ \hline
\hspace{0.5cm} MM Multi-Style 									& 8.58 	& 16.84 &  -8.27 & 4.97 & .23 & . & .79 & 132.04 & 200. \\ \hline
Fixed Income 																		& 9.46 	& 12.61 &  -3.15 & 2.57 & .30 & .19 & .60 & 183.83 & 183.23 \\ \hline
\hspace{0.5cm} FI Macro/Directional 						& 11.01 & 7.63 	&   3.38 & 5.67 & .24 & .20 & .69 & 127.6 & 88.35 \\ \hline
\hspace{0.5cm} FI Relative value 								& 9.17 	& 11.87 &  -2.70 & 2.33 & .20 & .08 & .66 & 19.78 & 27.07 \\ \hline
\hspace{0.5cm} FI Multi-Sector 									& 9.64 	& 8.36 	&   1.29 & 3.41 & .21 & .12 & .73 & 7.30 & 9.21 \\ \hline
\hspace{1.0cm} FI Government 										& 8.96 	& 7.27 	&   1.69 & 3.69 & .22 & .20 & .71 & 4.71 & 4.42 \\ \hline
\hspace{1.0cm} FI Corporate 										& 10.61 & 13.29 &  -2.67 & 4.30 & .49 & .32 & .44 & 8.57 & 9.24 \\ \hline
\hspace{1.0cm} FI Mortgages/Securitized 				& 9.02 	& 9.10 	&   -.08 & 2.82 & .21 & .27 & .66 & 3.02 & 3.92 \\ \hline
Equity 																					& 8.33 	& 19.67 & -11.34 & 7.19 & .22 & .03 & .77 & 6.30 & 8.40 \\ \hline
\hspace{0.5cm} EQ Macro/Directional 						& 8.43 	& 24.25 & -15.83 & 8.06 & .15 & .02 & .82 & 5.87 & 7.43 \\ \hline
\hspace{1.0cm} EQ Long-Short 										& 9.10 	& 23.22 & -14.12 & 7.91 & .22 & .05 & .75 & 7.24 & 9.71 \\ \hline
\hspace{1.5cm} EQ Long Only 										& 16.49 & 28.49 & -12.00 & 15.65 & .22 & .15 & .71 & 14.36 & 200. \\ \hline
\hspace{1.5cm} EQ Long Bias 										& 8.76 	& 21.49 & -12.74 & 9.49 & .17 & . & .84 & 7.30 & 8.37 \\ \hline
\hspace{1.5cm} EQ Variable Bias 								& 8.67 	& 20.81 & -12.14 & 6.07 & .21 & .11 & .74 & 11.65 & 17.54 \\ \hline
\hspace{1.5cm} EQ Short Bias 										& -.88 	& -12.03 & 11.14 & 9.80 & .06 & .06 & .90 & 164.82 & 90.8 \\ \hline
\hspace{0.5cm} EQ Relative value 								& 6.57 	& 12.04 &  -5.47 & 4.22 & .31 & . & .79 & 46.79 & 200. \\ \hline
\hspace{1.0cm} EQ Market Neutral 								& 5.33 &  7.37 &  -2.04 &  3.23 & .18 & .20 & .73 &  4.85 & 5.99 \\ \hline
\hspace{1.0cm} EQ Other Relative Value 					& 7.88 	& 14.67 &  -6.79 & 5.47 & .25 & . & .83 & 15.31 & 38.23 \\ \hline
\hspace{0.5cm} EQ Multi-Sector 									& 8.72 	& 20.08 & -11.36 & 6.67 & .30 & .08 & .65 & 9.70 & 15.61 \\ \hline
\hspace{1.0cm} EQ Energy 												& 10.43 & 23.90 & -13.47 & 10.95 & .21 & .09 & .82 & 7.57 & 10.88 \\ \hline
\hspace{1.0cm} EQ Finance 											& 11.01 & 3.18 	&   7.83 & 7.60 & .12 & .05 & .84 & 46.04 & 200. \\ \hline
\hspace{1.0cm} EQ Healthcare 										& 9.88 	& 9.60 	&    .29 & 16.09 & .20 & .25 & .67 & 5.67 & 7.01 \\ \hline
\hspace{1.0cm} EQ Technology 										& 4.38 	& 18.17 & -13.79 & 11.55 & .10 & .12 & .87 & 5.32 & 6.28 \\ \hline
\hspace{1.0cm} EQ Real Estate 									& 7.20 	& 12.61 &  -5.41 & 8.27 & .42 & .16 & .55 & 3.87 & 4.26 \\ \hline
FX 																							& 5.62 	& 2.81 	&   2.81 & 7.35 & .10 & .08 & .85 & 200. & 28.51 \\ \hline
Commodities 																		& 8.20  & -3.58 &  11.78 & 11.92 & .17 & .08 & .60 & 191.59 & 60.7 \\ \hline
Convertibles 																		& 7.88 	& 12.13 &  -4.25 & 4.35 & .28 & .07 & .75 & 5.02 & 7.01 \\ \hline
Broad Market 																		& 8.28 	& 16.56 &  -8.28 & 5.75 & .24 & .01 & .61 & 200. & 122.73 \\ \hline
\hspace{0.5cm} BR Developed Only 								& 7.25 	& 13.13 &  -5.87 & 4.88 & .25 & . & .80 & 8.54 & 10.26 \\ \hline
\hspace{0.5cm} BR Emerging Only 								& 18.18 & 35.33 & -17.14 & 10.9 & .22 & .31 & .58 & 11.31 & 29.12 \\ \hline

\end{tabular}
	\caption{Estimation of GAARCH model for LBHF asset class subindexes}
	\label{berd:tab:asset}
\end{table}

\subsection{Size Dependence} \label{berd:sec:size}

As we can see from Table \ref{berd:tab:size}, the model fit is fairly uniform across all the size subindexes, with the unconditional volatility gradually decreasing from 5.30\% across all funds, to 4.31\% across funds with AUM greater than \$1 billion. Note that the distribution of fund sizes is heavily dominated by smaller funds, and therefore the equal-weighted indexes with low size cutoff are actually representative of small fund performance.

The most notable size dependence is the gradual convergence of the returns to more symmetric and normal distribution as the cutoff fund size grows. This is evidenced both by the diminishing difference between the downside ($\eta^{-}$) and upside ($\eta^{+}$) ARCH coefficients and by the growing degrees of freedom both in the left ($\nu^{-}$) and the right ($\nu^{+}$) tails. Moreover, for the funds greater than \$500MM the right tail of the distribution actually becomes normal ($\nu = 200$ is a limiting value in our numerical estimation procedure, indicating a convergence to normal case which formally corresponds to $\nu \rightarrow +\infty$). 

This could be explained by the hypothesis that the larger funds tend to be more internally diversified, following multi-strategy investment process, and are therefore, on average, less subject to tail risk and asymmetric leverage. It could also be a sign of a more stringent risk management and especially more careful operational risk management in larger funds, which minimizes the likelihood of margin call induced forced selling and subsequent drawdowns (notwithstanding the negative examples of Amaranth, Sowood and few other funds).

Another visible dependence is the growing importance of the time-variability of volatility for bigger funds, evidenced by the relatively larger values of the ARCH coefficients and smaller values of the GARCH coefficient. This could be simply due to the fact that we have many fewer funds in the large fund category, and therefore the subindex itself is much less diversified, leading to greater importance of idiosyncratic return shocks month to month.

An important caveat is that the number of funds drops significantly as the cutoff size grows. Therefore, our results for the larger size funds are subject to greater statistical errors than for the smaller ones. 

\subsection{Style Dependence} \label{berd:sec:style}

The results of the model fit for different style subindexes of the LBHF Index are shown in Table \ref{berd:tab:style}. Here, there is a far bigger diversity of results, stemming from significant differences in the investment processes followed by the managers in each of these peer groups. The subindexes are grouped in three major sectors: Macro/Directional, CTA, and Relative Value (RV), with finer subdivisions within each sector. We see dramatic contrast between the groups with respect to asymmetry and fat-tailedness of returns. 

For example, some strategies, such as Event Driven, Distressed and Broad Relative Value, and all of the CTA strategies, have tails approaching the normality, as evidenced by large values of tail degrees of freedom parameters. On the other hand, certain strategies, such as the Macro, Long-Short with Variable Bias, and many of the Relative Value strategies, exhibit strong fat tails in their periodic returns distribution.

The static asymmetry of periodic returns is captured by the difference in the left and right tail parameter. We see some cases where this asymmetry is very large, e.g.\ LS Long Only strategies, which has fat left tail and normal right tail, CTA composite, which has a normal left tail and a mildly-fat right tail, the Relative Value composite which has a mildly-fat left tail and a normal right tail. 

The dynamic asymmetry of returns, captured by the difference between the downside and upside ARCH coefficients, is also markedly different across the strategy styles. For some, such as Macro composite, CTA composite and CTA Trend-following, LS Long Only, LS Short Bias, RV Merger Arbitrage, and RV Market-Neutral Equity, the difference is minimal and the dynamic asymmetry is insignificant. For others, such as Fundamental CTA, and vast majority of the relative value strategies, the downside ARCH coefficient is much greater than the upside one, signifying that there is a strong dynamic asymmetry, which is known to generate strong downside asymmetry of aggregate returns over long horizons (see \cite{Berd-Engle-Voronov-2007} and references therein). There are no strategies where the upside ARCH coefficients would be much greater than the downside, i.e.\ there are no strategies which would exhibit strong upside asymmetry of the aggregated returns over long horizons.

In each of these cases, one can see the vestiges of the corresponding strategy process in the distribution of its returns. Let's list some of the more obvious ones.

The CTA strategies trade futures, which have linear risk profile. Moreover, in vast majority of cases, they tend to have a reasonably low turnover, such that the biggest positions are held for a month or more. This makes the strategy returns over a month horizon (which is the periodicity of the returns in our sample) to be close to normal, because the daily fluctuations of the asset returns are allowed to aggregate over a month without strongly correlated position changes. This explanation is especially true for Fundamental CTA strategies, and less so for Trend-following ones which occasionally dabble in higher frequency trading that can scramble returns away from normal. Moreover, the Fundamental CTA, being more prone to holding large futures positions over long periods of time, exhibits a stronger dynamic asymmetry of returns which is more in line with general market behavior, whereas the Trend-following CTA has less dynamic asymmetry because its typical holdings are less biased towards being long the market.

On the other hand, most Relative Value strategies actually behave like a carry trade, selling something that is rich (lower yielding in expected return) and buying something that is cheap (higher yielding in expected return). For any such strategy, the typical return profile is highly asymmetric, because it can get caught in a forced unwind of this `carry trade', which usually happen violently. The proverbial `collecting nickels in front of a steamroller' strategy, which the Relative Value is in large part, is indeed subject to an occasional run-down by the steamroller! Hence, we see it manifested in both static and dynamic asymmetry of returns on the downside.

The few exceptions from this rule are also telling. Both the Event Driven and especially Distressed relative value strategies are quite different, in that they typically have much more balanced upside vs.\ downside trade profiles. For example, a distressed bond, which trades at 40 cents on a dollar, has an upside of 60 cents vs a downside of 40 cents, which is a dramatically different profile than a typical yield-based relative value bond trade, where one hopes to collect a few tens of basis points but takes a full principal risk (albeit with low probability). 

Finally, we can make sense of the patterns observed in various flavors of Long-Short strategies, by noting that the Long Bias  and Variable Bias strategies are similar to more imprecise and losely built relative value strategies, and correspondingly they exhibit similar features, namely fat tails and dynamic asymmetry of volatility. On the other hand, the Long Only and Short Bias strategies are quite different in their process. They are often run by fundamental managers who put on large conviction trades and do not rely on formal hedging for risk management. Most importantly, because they do not hedge, they also cannot employ a lot of leverage, which is usually the main cause of the asymmetric volatility response. This is indeed reflected in the much more symmetric pattern across the downside and upside fitted ARCH coefficients for these strategies.

\subsection{Asset Class Dependence} \label{berd:sec:asset}

The results of the model fit for different asset-class subindexes of the LBHF Index are shown in Table \ref{berd:tab:asset}. The segmentation by traded asset class appears to have less discriminatory power over the return characteristics, in part because within each asset class different hedge funds pursue different investment styles. In fact, the subsectors classified by the investment style exhibit a stronger dispersion of model characteristics, then the top level of this classification.

The most notable fact is that for those asset classes which admit a large variety of investment styles, encompassing both the Macro or CTA styles and the Relative Value styles, the corresponding top-level composite index exhibits almost normal distribution of periodic returns. This is most likely due to the mutually diversifying effects of those sub-styles in each asset class composite. The exceptions, Convertibles, and Equity composites, is most likely too dominated by thematic and sectoral funds and the corresponding composite is simply not diverse enough in style to achieve the normality of returns.

In this classification, we see for the first time the difference between the funds focused on the Emerging Markets vs.\ the ones trading in the Developed Markets only. It is telling that the Emerging Market funds exhibit larger variability of volatility, but with less asymmetry -- a feature generally present in strategies which are exposed to a lot of idiosyncratic return sources, as opposed to the ones whose returns are more systematic in nature. The lower typical leverage levels could also be an explanatory factor for this difference.

\section{The Nature of Alpha} \label{berd:sec:alpha}

Let us now turn our attention away from the volatility dynamics and distribution of return shocks and towards the expected returns and their dependence on the conditional volatility. In all three results Tables \ref{berd:tab:size},\ref{berd:tab:style},\ref{berd:tab:asset} we show the estimates of the true alpha $\alpha$, convexity compensation $\Gamma =  \gamma \sigma_0^2$ and the risk adjusted alpha $\hat{\alpha} = \alpha + \Gamma$.

\begin{figure}[t]
\includegraphics[height=3in,width=4.5in]{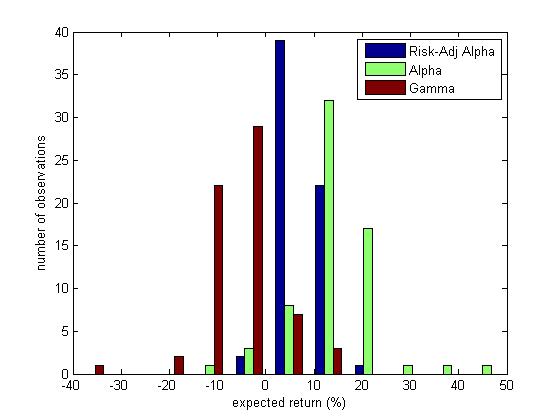}%
\caption{{\small Histogram of estimated values for risk-adjusted alpha, true alpha, and convexity compensation (gamma)}}%
\label{berd:fig:alphahist}%
\end{figure}

The first thing that strikes the eye is the remarkable uniformity of the risk adjusted alpha estimates across all LBHF subindexes, despite a substantially varied levels of true alpha and convexity compensation. The histograms of distributions of these model parameters are shown in Figure \ref{berd:fig:alphahist}. 

The few notable outliers are: LS Long Only (in style subindexes) and EQ Long Only (in asset class subindexes), which benefited from the bubble-esque runup in asset prices from 2000 until first half of 2008 (our sample period), and LS Short Bias and EQ Short Bias, which suffered from the same historical fact. We also note the outsized risk-adjusted returns in the Broad Emerging Markets subindex, which is explained by a secular rise in the emerging markets integration in the global capital markets and also to some extent by the super-fast economic growth and the bubble-like runup in the key commodity prices, affecting EM investments. 

Save for these few exceptions, the rest of the sectors exhibit risk-adjusted alpha between 5\% and 10\%, with most clustered around the magic `8\%' number that must warm the hearts of pension fund managers everywhere, because it is precisely what they often assume for the long-term expected returns in their portfolios. We will reserve the judgment on whether this number is reliable going forward, which is the subject of much discussion these days between the `Old Normal' and 'New Normal' paradigm proponents.

What is even more remarkable than the uniformity of the risk-adjusted alphas, is that the vast majority of hedge fund sectors exhibit positive true alpha, but negative convexity compensation $\Gamma$. Remembering that the same value of $\Gamma$ also enters the time-varying component of the conditional mean $\Gamma \chi_{t}$ (see eq.\ (\ref{gaarch_cmean_parts})), we are led to conclusion that when the volatility of the strategy returns increases ($\chi > 0$), these strategies tend to lose money, and vice versa, their returns get a positive boost whenever their return volatility is lower than the unconditional forecast ($\chi < 0$). 

It is important to note, that in our GAARCH model's conditional mean specification (\ref{gaarch_cmean}) the relationship is between the {\em forecast} of the conditional variance for the next time period $\sigma_{t+1}^2$, and the conditional expected (mean) return for that future period $\mu_{t+1}$. In other words this is not the usual leverage effect in financial time series, which runs in the opposite direction -- large realized returns precede increasing volatility. The effect we are describing might be consistent with a notion that somehow the relationship between the mean return and the scale of the return distribution in a given period is fixed, and then the predictive power of the volatility forecast over the mean return is simply by translation, due to volatility clustering. That would be an easy answer, except it would require that the mean return be always {\em positively} proportional to the forecasted volatility, which is actually the opposite of what we observe in the vast majority of cases.

\begin{figure}[t]
\includegraphics[height=3in,width=4.5in]{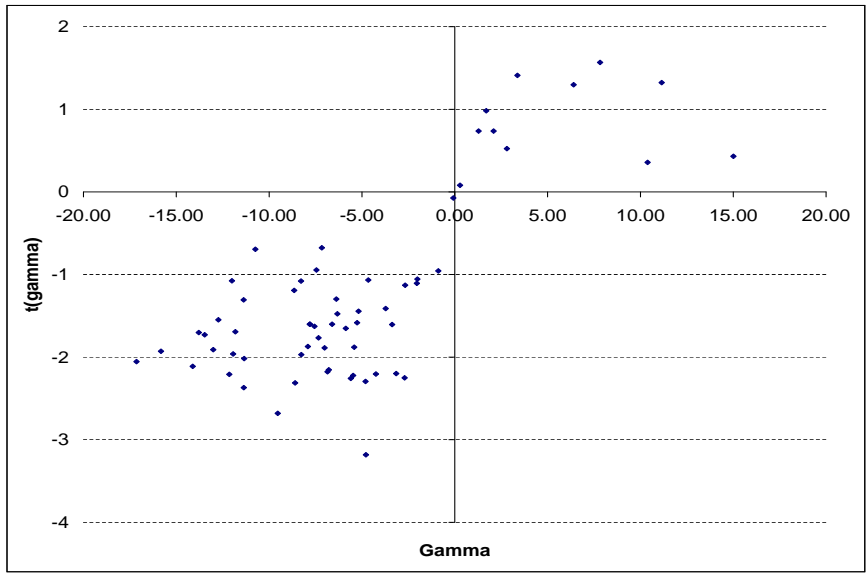}%
\caption{{\small Estimated t-statistics of the volatility exposure of conditional mean returns.}}%
\label{berd:fig:tstats}%
\end{figure}

In fact, we have discovered empirically that most hedge fund strategies have expected returns which are {\em negatively} proportional to the forecasted variance, and so are essentially {\em short vol}. Of course, this is not the same meaning of this phrase as usually used by traders, i.e.\ vast majority of these strategies don't actually have a negative exposure to some option-like instruments. Rather, what they have is a systematic pattern of investment which produces returns similar to those that would be produced by a short option position. Figure \ref{berd:fig:tstats} demonstrates that in most cases the volatility exposure coefficients are statistically significant. In fact, for most of the negative values observed we have t-stats greater than 1 in absolute value, and for some even greater than 2.

The statement about being short vol applies to all of the broad-based LBHF subindexes, all Relative Value subindexes, and to most Long-Short strategies, but it {\em does not apply} to LS Short Bias, CTA, CTA Trend-following style subindexes, and to FI Macro/Directional, FI Multi-Sector, FI Government, EQ Short Bias, FX and Commodities asset class subindexes. 

We believe that both the positive and the negative examples of this statement are very relevant for understanding of the nature of investment strategy alpha. In the remainder of this Section we will present a laundry list of both fundamental and technical reasons explaining why are most investment strategies short volatility, and what drives the exceptions.

\subsubsection*{Fundamental Reasons}
\begin{itemize}
\item Volatility is a synonym for `risk'. It is tough to both have a positive expected return {\em and} benefit from risk. If it was easily done, everyone would do it, and would thus change the properties of those strategies. So, the strategies appear to be short vol partly for Darwinian reasons, because they are successful.
\item Any well-defined investment strategy is intrinsically short an opportunity to reallocate to other strategies, which is usually triggered by either large loss or a large gain over some medium-term horizon. This can be interpreted as if the strategy contained a short straddle option on its own medium-term trailing returns, which naturally leads to a negative relationship between past volatility and future expected returns.
\item As a corollary to the previous point, we should mention that the cash asset can be considered a long opportunity option. The dry powder in your wallet gives you the opportunity to deploy it whenever you wish. Again, if the cash is long vol, then having spent the cash and invested in a strategy, the investors get shorter vol.
\item The ultimate demand to be long risk comes from the real economy, where the main risk takers, the entrepreneurs would like to share these risk exposures with outside investors. Therefore, investors in aggregate must be short the risk level (i.e.\ lose money when the risk goes up), otherwise they would serve no useful purpose for the real economy.
\end{itemize}

\subsubsection*{Technical Reasons}
\begin{itemize}
\item Liquidity reasons -- it is difficult to maintain a positive vol/convexity strategy with large capacity, unless one takes large directional risks {\em and} is willing to sacrifice potentially sizable carry costs. Providing liquidity is the only way to handle large capacity, and it leads to being short vol.
\item Leverage and financing -- most relative value strategies are designed to capture small pricing differences, and therefore must be leveraged to achieve reasonable nominal returns. This leverage and financing requirement, which is naturally short-term and subject to refinancing risk, exposes all relative value strategies to volatility risk (i.e.\ risk of uncertain volatility), when an increase in perceived vol will lead to higher margin requirement and trigger more trading and increased losses. This mechanism, in particular, was in abundant display during the recent crisis.
\item Any convergence, contrarian or relative value strategy that is betting on some sort of mean reversion or a convergence of some risk/return metrics is naturally short vol, because it explicitly bets against the wings of the returns distribution. We shall illustrate this point below in greater detail.
\item Many relative value strategies are really `carry trades', buying the higher yielding (cheap) asset and selling short the lower yielding (rich) one. And, like all carry trades, they are also naturally short vol because they are exposed to decompression risk, which is in turn directional with vol.
\end{itemize}

\subsubsection*{Reasons Behind the Exceptions}
\begin{itemize}
\item As a corollary to our statement about the convergence strategies, we can deduce that the momentum, trend-following strategies (which are essentially the opposites of the convergence) are naturally long vol. The CTA, CTA Trend-following, and Commodities strategies fall under this rule.
\item The niche for opportunistic/directional players always exists, but will always be limited in capacity. Notable exceptions from this are the directional fixed income (particularly in the government sector) and FX strategies, where the `other side' of the trade is the global economy and central banks, which are acting as risk absorbers rather than risk demanders. The FI Macro/Directional, FI Multi-Sector, FI Government, and FX strategies fall under this rule.
\item The Short Bias strategy's long vol exposure is somewhat coincidental -- the higher volatility is generally associated with downside markets, which in turn benefit the short bias strategy, so it is the market asymmetry that plays the role here.
\end{itemize}

\subsubsection*{Why are Convergence Strategies Short Vol, while Momentum Strategies are Long Vol?}

Any convergence strategy, when implemented in a disciplined way, will ultimately buy some assets when their price is `too low' and will sell them when their price is `too high', whatever the metrics they use to determine these thresholds. As a result, such strategies are betting that the prices will stay closer to the middle of the distribution and therefore betting against the wings of distribution. This is essentially a short strangle position on the assets, and is therefore naturally short volatility, but it benefits from a positive carry (collecting option premiums). 

\begin{figure}[t]
\includegraphics[height=3in,width=4.5in]{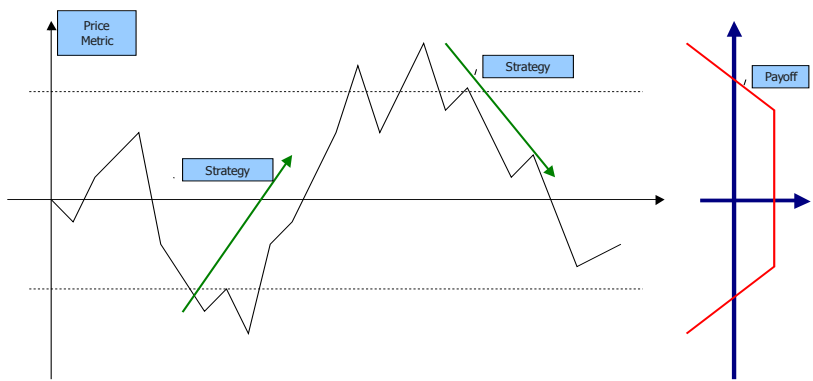}%
\caption{{\small Convergence strategies are short vol.}}%
\label{berd:fig:convergence1}%
\end{figure}

\begin{figure}[t]
\includegraphics[height=3in,width=4.5in]{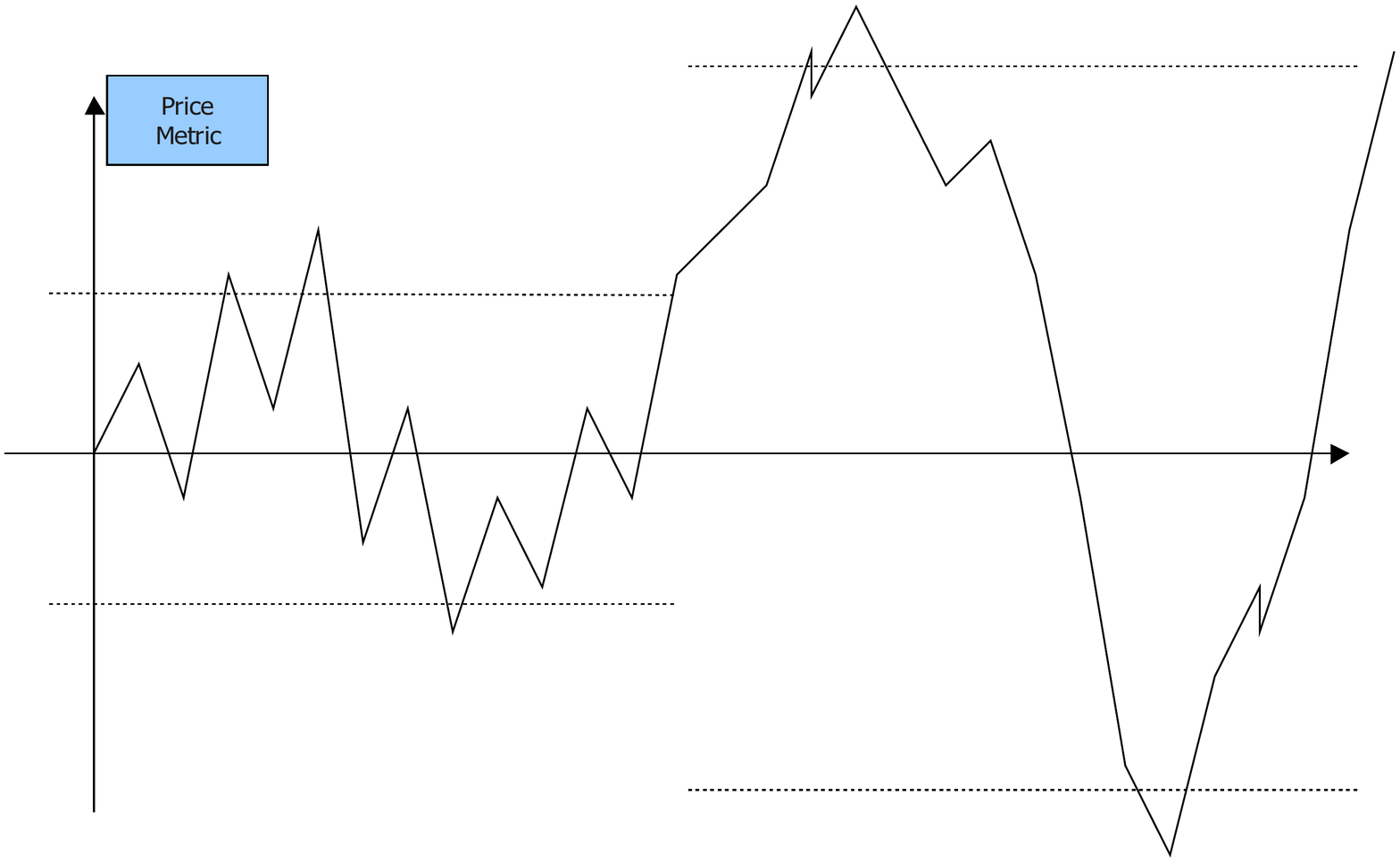}%
\caption{{\small Convergence strategies loose when vol goes up.}}%
\label{berd:fig:convergence2}%
\end{figure}

The propensity of convergence strategies to be short vol is even clearer at the times of sudden change in volatility regime. Even if the strategy was correct in both regimes, meaning that it has set the appropriate thresholds and will on average make money by buying low and selling high in each of these regimes, it will inevitably loose money when the low vol regime is transitioning to a high vol regime, and correspondingly the thresholds are being widened out. This logic is illustrated in Figures \ref{berd:fig:convergence1} and \ref{berd:fig:convergence2}.

\begin{figure}[t]
\includegraphics[height=3in,width=4.5in]{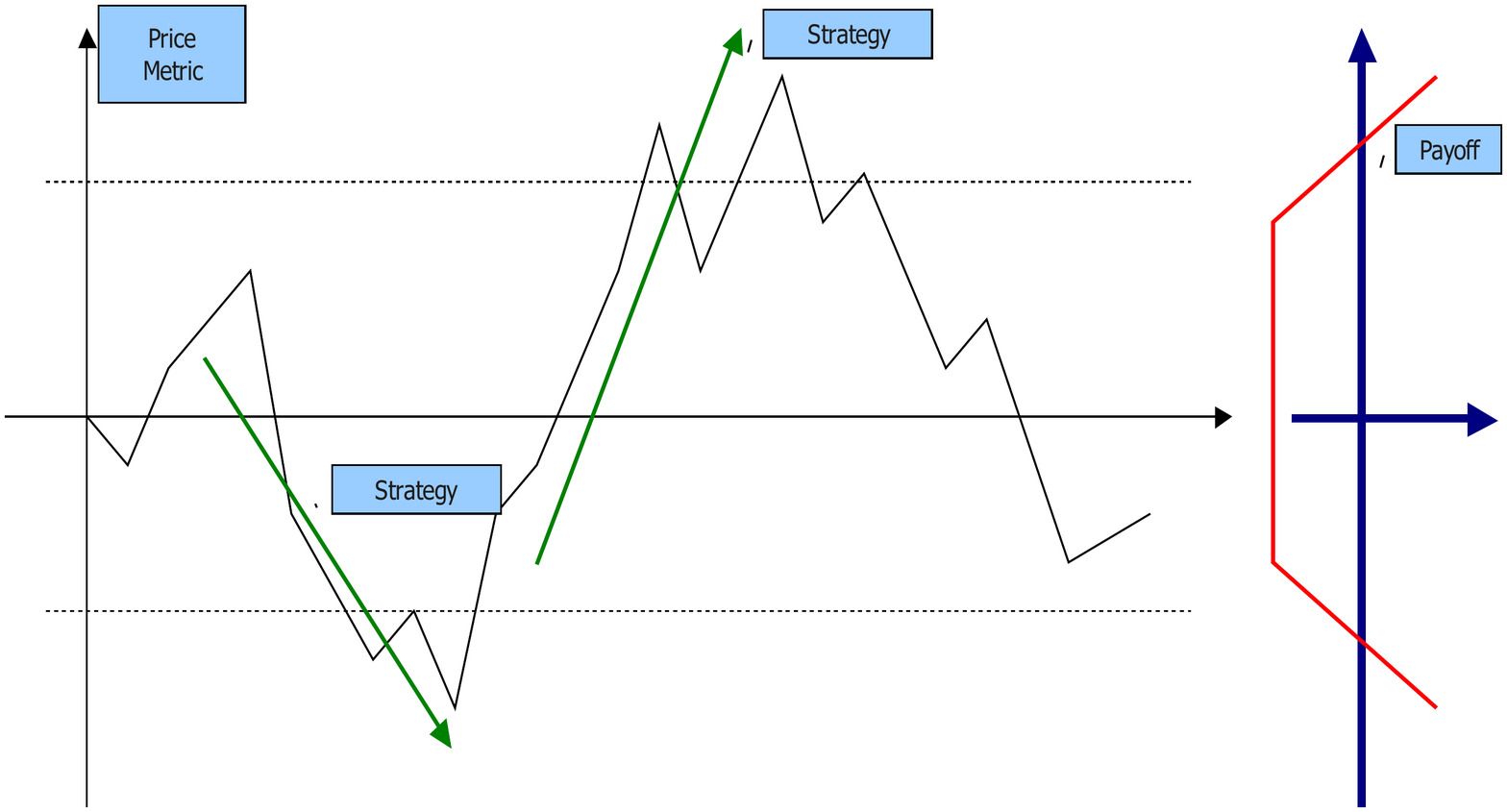}%
\caption{{\small Momentum strategies are long vol.}}%
\label{berd:fig:momentum}%
\end{figure}

On the other hand, any momentum strategy, when implemented in a disciplined way, will buy some assets when their price is `going up' and sell them when their price is `going down', using some sort of metrics to establish a proper time scale and thresholds. As a result, contrary to convergence strategies, these ones are betting that the prices will move further away from the middle of the distribution. This is essentially a long strangle position on the assets, and is therefore naturally long vol, but suffers from a short carry. This logic is illustrated in Figure \ref{berd:fig:momentum}. 

The observation that the trend-following strategies have option-like behavior has been made in a much more detailed manner in a notable paper by Fung and Hsieh \cite{Fung-Hsieh-2001}, building on an earlier work by Glosten and Jagannathan \cite{Glosten-Jagannathan-1994}. What we argue here, is that this insight applies almost uniformly across all hedge fund styles, once we recognize the differences in their volatility dynamics.

Note, that when we talk about the disciplined investment process, we do not necessarily mean a black-box, quantitative approach to investing. Certainly, it fits the bill. But so does also a process followed by most traditional, discretionary portfolio managers who have their own step-by-step approach to determining good investment opportunities. Warren Buffett is certainly not a quant, but his value style of investment is eminently disciplined and perhaps even repeatable, and is therefore very much subject to the same patterns that we have identified. In fact, one could probably explain a good deal of Buffett's investment returns by noting that he runs a distressed contrarian strategy (therefore somewhat short vol, but not as much as other relative value strategies, and less subject to asymmetric drawdowns), and he also keeps around a great deal of cash which, as we mentioned above, has some long vol characteristics, especially when coming out of recessions when the opportunity value of cash is the greatest.

\section{Practical Lessons} \label{berd:sec:lessons}

The most important result which we obtained here is that the vast majority of mainstream investment strategies have negative volatility exposure, whether or not they actually know it. We call this an {\em implicit volatility exposure}, to distinguish it from {\em implied volatility exposure} which one obtains when trading options. 

It is very difficult to manage such an exposure if the portfolio manager does not recognize it explicitly. The false sense of being hedged or being market-neutral lulls many portfolio managers into the trap of high leverage, and leaves them exposed to the risk of sudden margin calls when the vol goes up, which is the most painful way to discover one's dependence on a risk factor. 

We believe that this was in some way a contributing factor behind the Quant Crunch of 2007, as well as behind the more benign but equally befuddling behavior of quant strategies in 2009 after the beginning of the Fed's quantitative easing program. Just a month before the Quant Crunch, two Bear Stearns hedge funds focused on structured credit blew up, followed by Sowood, another credit hedge fund focused on leveraged loan investing. That marked the turning point for the market volatility which went up from the multi-year lows. As was argued in \cite{Khandani-Lo-2007}, the actual Quant Crunch occurred due to a liquidity squeeze in the crowded quantitative long/short equity space. But what precipitated this squeeze was the notch up in the volatility. 

As to the 2009 `great recovery' which left behind many traditional quant managers, it also seems to fit our description. What the Fed has done starting from March 2009, is acting as a risk absorber, and actively damping the level of all risks in the market. This leads to a fast drop in vol, and correspondingly large outperformance of convergence strategies. However, the outperformance was even greater for the extreme mean-reversion strategy pushing up the prices of every stock that went particularly far down in the preceding crisis period (what the CNBC pundits were calling a `melt-up' of the markets). This left the more mainstream slow-turnover quant strategies, which often identify risky stocks to short in order to maintain market-neutrality, dangerously squeezed by this run-up in the `crappy' stocks.

So, what should the investors do with all these insights? There are several ways one can manage the implicit volatility risks, once they have been identified. First and foremost, one can employ style and strategy diversification. Combining the strategies which are naturally short vol with those that are long will lead to a better overall portfolio. This is why all funds of funds must have a healthy allocation to CTAs, and other trend-following strategies, and to macro fixed income and FX strategies, since those are the ones that have positive vol exposure yet still can offer a reasonable capacity.

An even better approach is to add explicit (or implied) volatility exposure to the portfolio, by investing in volatility strategies. Volatility as an asset class has been on the rise in the recent years, although it still suffers from the lack of respect resulting in the absence of appropriate silo in which to put an allocation to this strategy (that is to say, it has not yet achieved a mainstream status) and many institutional investors do not think it can handle the size issues very well. Also, a lot of investors associate volatility strategies with just protective put option buying (`black swan' insurance), the cost of which is often prohibitive. 

We actually prefer dynamic volatility strategies mixing both directional, long/short, relative value and tail risk protection styles, which can provide the right amount of protection while also earning some alpha along the way, rather than costing an arm and a leg as pure insurance strategies do. We believe that the ideal volatility strategy mix must achieve return characteristics combining small but positive true alpha with positive (and large) convexity compensation, with fatter upside tail of the return shocks and low dynamic asymmetry. This cannot be achieved by trading volatility in any single asset class, but it might be possible if one builds a comprehensive volatility strategy encompassing different asset classes and investment styles \cite{Berd-VolCycle-2007,Berd-VolStrategies-2009}.



I would like to conclude by expressing my thanks to many colleagues with whom I discussed these topics over the years, including Robert Engle, Artem Voronov, Bob Litterman, Peter Zangari, Emanuel Derman, Mark Howard, Arthur Tetyevsky, Eric Rosenfeld, Marc Potters and Jean-Philippe Bouchaud.

\end{document}